# Exploring Quantum-Enhanced Machine Learning for Computer Vision: Applications and Insights on Noisy Intermediate-Scale Quantum Devices

**Purnachandra Mandadapu**
mpchandra39@gmail.com
Deloitte, Dallas, Texas, USA

*Abstract*— **As medium-scale quantum computers progress, the application of quantum algorithms across diverse fields like simulating physical systems, chemistry, optimization, and cryptography becomes more prevalent. However, these quantum computers, known as Noisy Intermediate Scale Quantum (NISQ), are susceptible to noise, prompting the search for applications that can capitalize on quantum advantage without extensive error correction procedures. Since, Machine Learning (ML), particularly Deep Learning (DL), faces challenges due to resource-intensive training and algorithmic opacity. Therefore, this study explores the intersection of quantum computing and ML, focusing on computer vision tasks. Specifically, it evaluates the effectiveness of hybrid quantum-classical algorithms, such as the data re-uploading scheme and the patch Generative Adversarial Networks (GAN) model, on small-scale quantum devices. Through practical implementation and testing, the study reveals comparable or superior performance of these algorithms compared to classical counterparts, highlighting the potential of leveraging quantum algorithms in ML tasks.**

*Index Terms*— **Deep Learning, Generative Adversarial Networks, Machine Learning, Noisy Intermediate Scale Quantum, Quantum Algorithms**

## I. INTRODUCTION

As medium-scale quantum computers continue to advance, the application of quantum algorithms becomes more prevalent across diverse fields such as simulating physical systems, chemistry, optimization, and cryptography. These algorithms exploit the distinctive properties of quantum mechanics, such as entanglement and superposition, to overcome the limitations encountered in classical computing, thereby achieving what is known as quantum advantage. At present, quantum computers typically operate within the range of small to medium scales and are susceptible to noise, a characteristic referred to as Noisy Intermediate Scale Quantum (NISQ). Consequently, there is a growing interest in identifying applications that can capitalize on quantum advantage without requiring error correction procedures, as implementing such routines would significantly escalate the demand for qubits. In simpler terms, as medium-sized quantum computers become more powerful, they're being used in various fields like simulating physical systems and encrypting data. These computers work differently from regular ones, using special properties of quantum mechanics.

In recent years, there has been significant growth in the field of Machine Learning (ML)—a problem-solving approach where machines learn to tackle tasks by digesting large volumes of data or general information pertaining to the problem [1]–[5]. This enables them to discern patterns and develop their own decision-making processes, without explicit instructions. Common applications include organizing data into categories, grouping similar items, generating new data, and facilitating autonomous driving systems. However, a notable challenge of this approach is the substantial need for data and computational resources for training, coupled with the inherent opacity of the algorithms, making it difficult to ensure optimal resource utilization. Given these challenges, it prompts the question: Can quantum computers offer assistance to ML algorithms? Specifically, can quantum properties enhance the effectiveness of ML processes without solely relying on faster hardware, which often poses a significant bottleneck in advancing ML research? Many studies have looked into the fascinating idea—using quantum technology to improve ML. Therefore, this study combines the strengths of both quantum computing and traditional methods. Quantum variational algorithms offer powerful problem-solving capabilities, while classical computing provides the infrastructure for large-scale optimization. Therefore, in this study, we focus on applying hybrid quantum-classical algorithms to a crucial area of ML—computer vision. Specifically, we examine their effectiveness in tasks like sorting images into categories and generating new images. Inspired by promising findings in existing literature, we implement and evaluate two such algorithms: the data re-uploading scheme and the patch Generative Adversarial Networks (GAN) model. These algorithms are tailored to work well even on small-scale quantum devices. Our experiments reveal that both the data re-uploading scheme and the patch GAN model perform comparably to or even better than classical algorithms of similar complexity. This underscores the potential of leveraging quantum algorithms, especially on smaller systems. It suggests that as quantum technology matures, we may see even greater advantages in the realm of ML.

The study commences by introducing fundamental concepts in quantum computation, including the notion of quantum circuits and quantum gates, elucidating their susceptibility to errors in Section II. These components serve as the foundational elements of quantum algorithms, particularly pivotal in the realm of Quantum Neural Networks (QNNs). Subsequently, an overview of related works is presented in Section III. Section IV brings together important concepts of quantum ML. We will define Variational Quantum Circuits (VQCs), which are quantum circuits with adjustable parts, and explore how they function as QNNs. Furthermore, we'll delve into two specific quantum ML algorithms—data re-uploading and Patch GAN. We'll explain these algorithms and provide reasoning for their utilization. In Section V, we will practically implement and test these algorithms. We'll conduct tests on both simulators, including perfect and noisy versions, and on a real quantum device for the data re-uploading algorithm. Finally, we'll draw conclusions based on our findings and outline potential future avenues for research in Section VI.

## II. BACKGROUND

In this section, we offer a brief introduction to quantum computation, amplifying the fundamental concepts important for this study. Quantum computation involves the processing and manipulation of data rooted in the principles of quantum mechanics, notably superposition and entanglement. Unlike classical computation, which uses the binary scalar bits for information representation—quantum computation utilizes quantum bits, or qubits, which are denoted by states within a Hilbert[1] space. Qubits are fundamental units in quantum computing, often represented using discrete-variable states such as $|0\rangle = \begin{pmatrix} 1 \\ 0 \end{pmatrix}$ and $|1\rangle = \begin{pmatrix} 0 \\ 1 \end{pmatrix}$. Unlike classical bits, qubits can exist in a state of superposition, expressed through changes in basis. This superposition is defined by $|+\rangle = \frac{1}{\sqrt{2}} \begin{pmatrix} 1 \\ 1 \end{pmatrix}$ and $|-\rangle = \frac{1}{\sqrt{2}} \begin{pmatrix} 1 \\ -1 \end{pmatrix}$, where $|+\rangle = \frac{1}{\sqrt{2}}(|0\rangle + |1\rangle)$ $and$ $|-\rangle = \frac{1}{\sqrt{2}}(|0\rangle - |1\rangle)$. Quantum gates, akin to classical logic gates, operate on qubits to preserve unitarity. One such gate is the Hadamard gate, denoted as $\hat{H}$, which transforms $|0\rangle$ $to$ $|+\rangle$ $and$ $|1\rangle$ $to$ $|-\rangle$. When dealing with multiple qubits, the Hilbert space is tensorized to accommodate entanglement between qubits. Therefore, in quantum mechanics, states such as those depicted in Equations (1-4) exemplify what is termed as the Bell[2] basis. These states represent a class of maximally entangled states, characterized by intricate quantum correlations between particles.

$$|\Phi^+\rangle = \frac{1}{\sqrt{2}}|00\rangle + \frac{1}{\sqrt{2}}|11\rangle \tag{1}$$

$$|\Phi^-\rangle = \frac{1}{\sqrt{2}}|00\rangle - \frac{1}{\sqrt{2}}|11\rangle \tag{2}$$

$$|\Psi^+\rangle = \frac{1}{\sqrt{2}}|01\rangle + \frac{1}{\sqrt{2}}|10\rangle \tag{3}$$

$$|\Psi^-\rangle = \frac{1}{\sqrt{2}}|01\rangle - \frac{1}{\sqrt{2}}|10\rangle \tag{4}$$

However, in quantum computing, a system of $n$ qubits is commonly represented by a grid structure known as a quantum circuit. This circuit visually displays the operations that manipulate the qubits within the Hilbert space. The Hilbert space, denoted as $H$, is the mathematical construct representing all possible states of the qubits. The operators, such as the $X$, $Z$, and $Y$, are depicted within the circuit, indicating the transformations applied to the qubits. For instance, the $X$ operator induces a bit-flip[3] operation, the $Z$ operator introduces a phase-flip[4] operation, and the $Y$ operator combines both bit-flip and phase-flip operations. This grid-based[5] representation serves as a foundational tool in understanding and analyzing quantum algorithms and their implementations. Quantum computing relies on fundamental building blocks called quantum gates. These gates perform operations on quantum bits, or qubits, which are the basic units of quantum information such as:

- **Hadamard Gates (H):** This gate is used to create superposition, a key principle in quantum computing where a qubit can exist in multiple states simultaneously.

- **Pauli Gates ($X$, $Y$, $Z$):** These gates are fundamental in quantum computing for manipulating qubits along different axes in the Bloch sphere, a geometric representation of qubit states.

- **Entangling Controlled 'Not' (CNOT) Gate:** This gate entangles two qubits, where the state of one qubit depends on the state of the other qubit. The CNOT gate applies the $X$ operator to the target qubit based on the value of the control qubit.

Quantum computing, although promising, faces challenges due to errors stemming from various sources such as environmental factors and inherent qubit imperfections. These errors are quantified using Kraus[6] operators, offering a formal framework to model noise in quantum circuits. The interaction between the quantum system and its environment is crucial, often resulting in undesired effects on the quantum information. One prevalent type of noise is the bit-flip channel, where qubits undergo unintended changes analogous to classical bit flips. Similarly, the phase flip channel introduces alterations akin to changes in the quantum phase. Moreover, the amplitude damping[7] channel represents a significant error source, mimicking energy dissipation in the quantum system due to environmental interactions. Another form of noise, the depolarizing[8] channel, adds complexity by introducing a combination of random gate errors. Understanding how quantum circuits behave under these errors is essential for assessing the practicality of quantum algorithms. Moving forward, the performance of quantum circuits, especially those used as neural networks, will be evaluated, with quantum noise incorporated into the circuits to simulate real-world conditions or inherent imperfections in quantum hardware.

---

[1] A term from mathematics that is used in functional analysis and quantum mechanics to describe an entire, infinite-dimensional vector space.
[2] A group of four maximally entangled states that serve as the foundation for Bell experiments' description of quantum correlations.

[3] A quantum gate that changes a qubit's state from $|0\rangle$ to $|1\rangle$ or the other way around.
[4] A quantum gate that reverses a qubit's phase, converting $|0\rangle$ to $|0\rangle$ and $|1\rangle$ to $-|1\rangle$.
[5] A technique used in geometric and computational techniques that discretizes space or data into a grid form.
[6] A family of linear operators that characterize open quantum system dynamics in terms of quantum channels.
[7] A quantum channel that simulates energy dissipation or decoherence by causing the amplitude of a qubit to decay.

## III. RELATED WORKS

Hybrid quantum-classical ML methods have recently attracted considerable attention due to their ability to combine the strengths of classical and quantum computing, offering potential solutions to computationally intractable problems with greater efficiency. These methods seek to utilize quantum computing for specific tasks while relying on classical computing for data preprocessing and result postprocessing. In a recent study [6], a hybrid quantum-classical Convolutional Neural Network (CNN) model was proposed for X-ray image prediction. The quantum component of the model involves encoding, random quantum circuits, and decoding phases. This hybrid model demonstrated superior accuracy, outperforming classical ML approaches in sensitivity and F1-measure. Another hybrid quantum-classical CNN model was introduced in a separate study [7], employing a federated learning approach to enhance model security and mitigate privacy attack vulnerabilities. Results indicated that models with quantum convolution exhibited slightly improved accuracy compared to baseline classical models. Additionally, researchers proposed a hybrid quantum-classical model of Long Short-Term Memory (LSTM), a type of Recurrent Neural Network (RNN), in another study [8]. Comparative analysis revealed that the hybrid model achieved faster convergence and higher accuracy than its classical counterpart, albeit under conditions assuming absence of noise and decoherence. Furthermore, a hybrid quantum-classical approach was developed for generative adversarial learning in [9], focusing on anomaly and fraudulent transaction detection. Performance evaluations showed comparable results to classical methods in terms of F1 score. Recent work [10] delved into catastrophic forgetting with quantum algorithms, particularly focusing on incremental learning across different classification tasks. Inspired by replay methods, the researchers proposed constraining model updates by projecting gradient directions onto regions defined by previous task gradients, along with storing a portion of training data from previous tasks for gradient descent computation. However, a drawback of this method lies in the need to compute gradients of previous tasks at each training iteration. Overall, recent advancements in hybrid quantum-classical ML methodologies underscore their potential to surpass classical algorithms, particularly in handling large and complex datasets. As quantum computing technology continues to advance, further progress and the development of more powerful hybrid algorithms are expected in this field.

## IV. MATERIALS AND METHODS

In recent years, researchers have been exploring the potential of quantum computers to tackle linear algebra problems [11]–[14]. While there have been efforts to demonstrate a quantum advantage in conjunction with ML algorithms [15], [16], [25]–[27], [17]–[24], these endeavours encounter common challenges—the practical limitations of the Harrow-Hassidim-Lloyd [9](HHL) algorithm once implemented, and the substantial resources, including qubits and circuit depth, that they demand. To comprehend the applicability of NISQ in the realm of ML, it becomes imperative to rely on quantum algorithms complemented by classical counterparts, particularly in scenarios where quantum algorithms face implementation barriers, such as solving optimization problems related to fitting functions. VQCs emerge as promising candidates for integrating quantum computation into ML tasks due to their small scale and relatively lower susceptibility to noise [28]. This makes them particularly suitable for current quantum devices' capabilities.

### A. Variational Quantum Circuits

VQCs are a type of quantum-classical architecture that blends quantum and classical computing components. A VQC comprises a parameterized quantum circuit, followed by a classical optimization routine like Stochastic Gradient Descent (SGD). Its design is straightforward, consisting mainly of two components—the quantum circuit and the classical optimizer. The quantum circuit is typically constructed using a series of quantum gates, often parameterized, such as single or controlled rotation gates. These gates manipulate quantum states, with the rotation angles serving as parameters adjusted by the classical optimizer during the optimization process. Meanwhile, the classical optimizer plays a crucial role in determining the optimal values for these parameters, aiming to minimize a specified cost function. These cost functions vary, being either problem-specific, such as those in molecular energy level studies [29], or problem-agnostic, typically associated with classical supervised learning scenarios [30]. Generally, the fundamental steps involved in a VQC are shown in Algorithm 1.

| Algorithm 1: VQC |
| --- |
| **START** |
| 1.   *Initialization of the quantum circuit with initial parameter values.* |
| 2.   *Setting the qubits to a defined state, often the $|0\rangle$ state.* |
| 3.   *Apply the parameterized quantum circuit to an input state.* |
| 4.   *Measuring the quantum circuit's output using predefined measurement operators.* |
| 5.   *Computing the cost function based on the measurement outcomes, which relies on the circuit's parameters.* |
| 6.   *Utilizing the classical optimizer to refine the quantum circuit's parameters.* |
| 7.   *Repeat steps 2-5 until convergence is achieved.* |
| **END** |

---

[8] Often employed in quantum error correction, a quantum channel randomly rotates qubit states, adding noise and decreasing fidelity.

[9] A quantum technique with potential applications in optimization that solves linear equation systems quickly.

A notable advantage of VQCs lies in their potential to solve problems more efficiently than classical algorithms. This efficacy stems from the capacity of quantum circuits to explore exponentially large quantum state spaces [31], facilitating more effective search for optimal solutions compared to classical approaches. When we use a VQC, we adjust some settings to get the best result. It's a bit like how a neural network works. Next, we want to see how we can combine the best parts of both quantum computing and neural networks. When a VQC gives us an result, it's usually in a special format. To work with this result, we need to be able to change the settings and see how the result changes. We have a way to do this called the parameter shift rule[10]. For instance, the rotation gate—can easily show how the result changes. We can use this to make our VQC work better. We can even use it to help us understand how different parts of the VQC affect the final result.

However, in VQCs, various types of noise and errors affect their performance. Unlike traditional algorithms with fixed parameters or 'black-box' gates, VQCs are learned during the optimization process, making it crucial to analyze how different errors impact their effectiveness. One prevalent type of noise is stochastic noise, which arises during the optimization process. However, the saddle points[11], where the first derivative is zero but the hessian[12] matrix has eigenvalues of different signs, pose challenges. To mitigate the risk of being trapped in saddle points, methods like the SGD have been proposed, which introduces stochastic variables into the update rule. Measurement noise is inherent in VQCs, particularly during the estimation of derivatives. This noise arises due to the stochastic nature of quantum measurement. By considering the difference between the actual derivative and the estimation, a SGD algorithm naturally emerges, without the need for additional random variables. VQCs exhibit robustness against certain types of noise, such as bit flip-like and phase flip-like errors, as discussed in [28]. These errors are suppressible if they preserve the circuit symmetry. However, decoherent noise, like amplitude damping and depolarizing channels, remains a significant challenge. Decoherent noise can lead to difficulties in executing quantum algorithms correctly, necessitating innovative approaches to mitigate its effects. Recent studies have explored methods to protect information contained in states, offering promising avenues for addressing decoherence errors in VQCs.

### B. Quantum Neural Networks

The QNNs operates using quantum circuits, replacing traditional neurons and layers with quantum components like qubits and quantum gates. QNNs process both quantum and classical data, with quantum simulation showing promise in solving complex problems in quantum systems, such as simulating molecular spectra or quantum dynamics. This approach is particularly relevant in fields like condensed matter and quantum chemistry due to their natural mathematical description using Hamiltonian[13] operators. QNNs have diverse applications in quantum simulation, including Hamiltonian interaction mapping, ground state compression, and exploring quantum materials and phase transitions. VQCs have shown promise in this domain, outperforming alternatives like the Quantum Phase Estimation (QPE) algorithm, which is susceptible to noise and errors. Experimental trials on IBM's Qiskit[14] framework using the ibmq-bogota[15] device have highlighted the challenges posed by noise in QPE and the robustness of VQCs in such environments (refer to Fig. 1). While quantum data analysis offers advantages in studying quantum states directly on a quantum device—the potential of quantum computing to enhance classical ML algorithms remains unexplored. An additional challenge in classical ML is the tendency of neural networks to over-parameterize, leading to difficulties in data generalization. Recent studies have focused on whether quantum circuits offer superior capabilities in sampling distributions and leveraging the high dimensionality of Hilbert space, where classical data must be embedded.

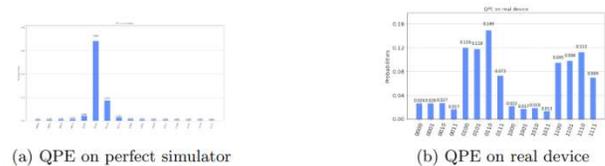

(a) QPE on perfect simulator    (b) QPE on noisy device

Fig. 1. QPE comparison between a ideal device and an noisy device

However, in Fig. 2, classical data is encoded into a quantum circuit using different methods known as ansatzes[16]. These include amplitude embedding, assigning data values to qubit amplitudes; basis embedding, suitable for binary data mapped to qubits in $\sigma_z$ base; and angle embedding, applying rotation matrices to encode data phases. Choosing the optimal ansatz is actively researched due to their distinct pros and cons. While amplitude embedding is straightforward, it requires careful selection of a small constant to ensure non-zero amplitudes for zero-valued data points, and basis embedding scales logarithmically with qubit number and is limited to binary data. Angle embedding requires an additional encoding gate, increasing circuit depth. Recent studies has also explored the inefficiency of certain ansatzes, particularly those resulting in product states like ($x = (x_1, x_2, x_3) \rightarrow |x_1\rangle|x_2\rangle|x_3\rangle$). This underscores the importance of selecting efficient encoding methods for quantum data analysis.

---

[10] A method that uses the difference of circuit outputs with shifted parameters to estimate gradients in quantum circuits.

[11] There are places in the optimization landscape where gradients disappear—they are neither minimum nor maximum—and this presents difficulties for optimization convergence.

[12] A square matrix of second-order partial derivatives that is employed to describe a scalar-valued function's curvature.

[13] In quantum physics, representations of a system's total energy are essential for comprehending its dynamics and long-term evolution.

[14] https://qiskit.org/index.html

[15] https://metriq.info/Method/97

[16] Forms or structures that are hypothesized and employed as first approximations for algorithms in problem solving, especially in quantum computing.

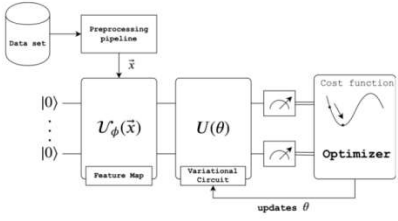

Fig. 2. Working blocks of a QNN

Recent studies have seen a surge in quantitative findings, often rooted in statistical learning theory. This field meticulously examines statistical models to quantify concepts like expressibility and generalizability. Expressibility refers to a model's ability to handle complex functions with limited data, while generalizability involves accurately predicting unseen data. Of particular interest is exploring the information-theory aspects of QNNs [32], focusing on analyzing the Fisher[17] information matrix. The Fisher information matrix is a vital tool for gauging a model's trainability, often mirroring the Hessian matrix of the loss for common loss functions. By analyzing its eigen values, researchers gain insights into loss function behavior, enabling them to anticipate barren plateaus that impede gradient descent optimization and lead to suboptimal models. Recent studies [32], [33] have delved into the effective dimension of models within their parameter space, highlighting a positive correlation between larger eigenvalues of the Fisher matrix and enhanced model efficacy, particularly evident in QNNs [32]. However, viewing neural networks through the statistical learning theory lens, they are seen as statistical models describing the relationship between data pairs $(x, y)$, where $p(x, y; \theta) = p(y \mid x; \theta) p(x)$. Here, $(x)$ represents input data points—features, $(y)$ represents output data points—labels for classification tasks or scalar values for regression tasks, and $(\theta)$ denotes the empirical parameter space. This construction explicitly accounts for a finite number of available data points.

However, Quantum Convolutional Neural Networks (QCNNs) have garnered interest for image processing akin to classical CNNs, but initial studies like [34] debunk the effectiveness of simple quantum feature maps. Subsequent efforts focus on integrating quantum operations with convolutional layers, inspired by classical Deep Learning (DL) architectures' modular design. This aims to create quantum computation modules that seamlessly merge with conventional layers, as illustrated in Fig. 3. In QCNNs, input data undergo processing through VQCs, potentially outperforming classical kernels. Quantum measurement aids in extracting pixel values and acts as a nonlinear activation function. Despite challenges, various QCNN approaches show promise. A compelling solution is the data re-uploading scheme—streamlining circuit components and enhancing efficiency. Its non-linear mapping from input to Hilbert space aligns with the universal approximation theorem, showcasing potential. Flexibility with multiple qubits enhances model expressibility. The scheme presents a promising avenue for advancing QCNNs, addressing both theoretical and practical aspects effectively. Similarly, the potential of VQCs as generative models has been thoroughly investigated recently, both in terms of theory and experimentation. High expressibility in describing continuous distributions of unstructured data is provided by VQCs. Their use as generators in the discriminator-generator configuration of GANs is one appropriate application. Patch GAN is the proposed approach to manage computational resources and optimize performance (refer to Fig. 4). The generator in a patch GAN is made up of several subcircuits, each of which is designed to produce a different percentage of the high-dimensional data. Each subcircuit uses controlled $Z$ gates for entanglement and trainable unitary single qubit rotation gates in response to an input noise vector. Partial measurements allow for the use of nonlinear mappings that improve model expressibility even more. Concatenating the outputs of each subgenerator yields the final output, which allows limited qubits to generate high-dimensional data. The potential of quantum computing in generative modeling is demonstrated by this approach, which uses $RY$ rotations and $CZ$ entangling gates to generate images on benchmark datasets like MNIST[18] and CIFAR10[19].

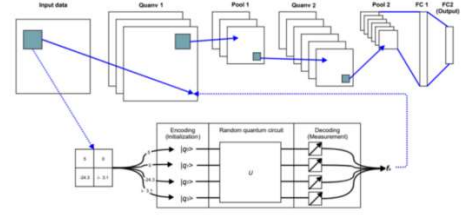

Fig. 3. QNN Schematization

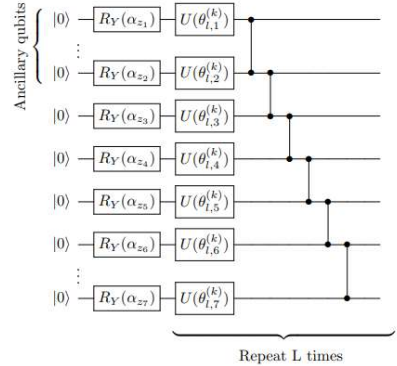

Fig. 4. Quantum circuit for patch quantum GAN

## V. RESULT ANALYSIS

To illustrate the impact of using quantum computation in ML tasks, practical applications on real-world tasks are essential. Among various fields of computer science, computer vision stands out as one heavily reliant on ML techniques, particularly in recent years, thanks to advancements in computational hardware and the refinement of DL models. The complexity of computer vision tasks arises from the high-dimensional nature of image data, where even a simple RGB image with $N$ by $N$ pixels translates into a tensor of dimensions $3N^2$. This complexity drives continuous efforts to enhance both computational power and algorithmic efficiency, making the integration of quantum computation into computer vision a promising prospect for the future. In adapting quantum ML algorithms, as discussed in Section 4, for computer vision tasks, adjustments are required, particularly concerning the data reuploading scheme. While classical CNNs process data locally, preserving spatial characteristics through convolutions, the data reuploading approach is tailored to handle localized segments of input data. The application of quantum ML in computer vision encompasses two primary areas—image generation and image classification. For image generation tasks, the effectiveness of the patch GAN architecture will be demonstrated using standard computer vision benchmark datasets, including MNIST, Fashion MNIST[20], and CIFAR10. Furthermore, beyond these standardized datasets, real-world challenges will be addressed, such as the identification of Parkinson's disease through the analysis of Positron Emission Tomography (PET) scan brain images [35]. The focus of this study lies in the thorough analysis of an algorithm, with particular attention paid to its performance and scalability. Additionally, the impact of noise on the outcomes will be carefully addressed. Subsequently, the results derived from the aforementioned architectures will be presented and juxtaposed against those obtained from classical neural networks possessing a comparable number of parameters.

---

[17] A matrix that shows the likelihood function's curvature in a statistical model, which is important for estimating parameters.

[18] http://yann.lecun.com/exdb/mnist/
[19] https://www.cs.toronto.edu/~kriz/cifar.html

## A. Image Classification

To align with CNNs, the approach of data reuploading ansatz has undergone adaptation. Instead of processing the entire image within the quantum circuit, which is computationally inefficient and time-consuming, a method akin to classical practices has been adopted. This involves applying a sliding window technique to the image. Rather than conducting convolution between the pixels under the window and the neural network's kernel, the pixel values within the window are inputted to the quantum circuit. The quantum circuit functions as a quantum kernel, following a similar framework as elucidated in [31]. The development of the QNNs—aims for seamless integration into the TensorFlow framework. These are implemented as callable Python classes, facilitated by the TensorFlow-Quantum[21] and Cirq[22] libraries. Importantly, all quantum layers are compatible with the classical pipeline of standard DL models, ensuring full interoperability with other layers present in the library. We investigate the scalability of the data reuploading scheme through adjustments in the number of layers and qubits. Utilizing the MNIST dataset, a compact QNNs architecture comprising two convolutional layers with an intermediate maxpool operation, followed by a fully connected classical softmax output layer, is evaluated to gauge its practicality. We focus on examining how changes in classification accuracy correlate with variations in the number of reuploading layers within each convolutional layer and the number of qubits employed in the quantum circuit. Furthermore, we conduct additional tests to explore the impact of noise within the quantum circuit, specifically assessing how classification accuracy is affected by varying amplitude damping parameters. The impact of amplitude damping noise on the effectiveness of quantum circuits has been observed to be quite significant. It's crucial to demonstrate the resilience of specific algorithms, even when faced with minor instances of amplitude damping, as this is essential in the quest for quantum advantage. To thoroughly analyze the ideal system, numerical simulations of quantum circuits were conducted. These simulations allowed us to focus solely on the effects of amplitude damping noise. This noise was introduced as a quantum operation, following Kraus operators, at the conclusion of each quantum circuit within the neural network, as illustrated in Fig. 5. Fig. 6(a) demonstrates that the inclusion of more reuploading layers indeed enhances performance, reinforcing the effectiveness of the reuploading technique. Furthermore, Fig. 6(b) illustrates the pivotal role played by the number of qubits in the quantum circuit's performance. It highlights how a quantum circuit with multiple qubits, entangled together, offers greater flexibility in terms of the functions it can accommodate.

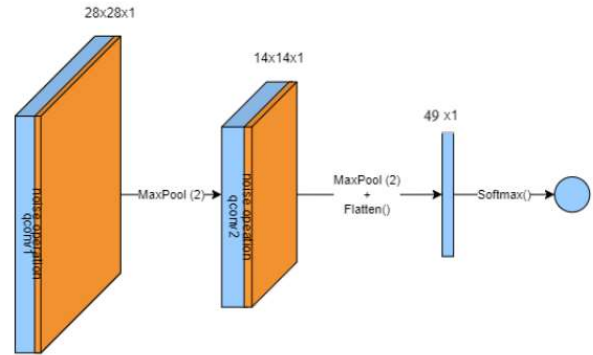

Fig. 5. The performance analysis in the presence of amplitude damping noise was conducted using QNN

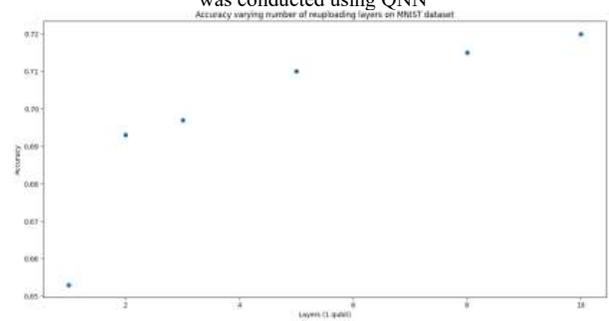

(a) Accuracy vs number of reuploading layers

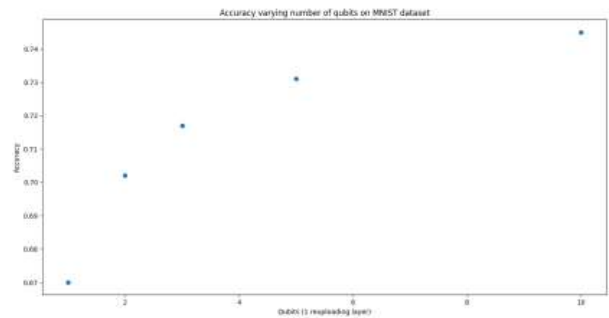

(b) Accuracy vs number of qubits

Fig. 6. Reuploading image classifier accuracy on the MNIST dataset with different reuploading layers/qubits compared to a constant number of layers/qubits

In this, the focus is on exploring the performance of QNNs in classifying various benchmark datasets. Specifically, the analysis involves comparing the QNN with classical CNN across different datasets such as MNIST, Fashion MNIST, CIFAR10, and brain PET images (CDOPA). The MNIST dataset comprises grayscale images depicting digits from 0 to 9, Fashion MNIST features clothing items, CIFAR10 consists of RGB images depicting various objects and animals, while CDOPA involves binary classification of PET brain images for diagnosing Parkinson's disease. For each dataset, the QNN is trained using an ideal simulator, a noisy simulator introducing bit and phase flip errors, and a real quantum computer, namely Google's IonQ Harmony[23]. The test accuracy and training-validation curves are analyzed and presented in Figs. 7, 8, and 9. The architecture of the neural networks, including the number of trainable parameters and hyperparameters, is detailed in Table I.

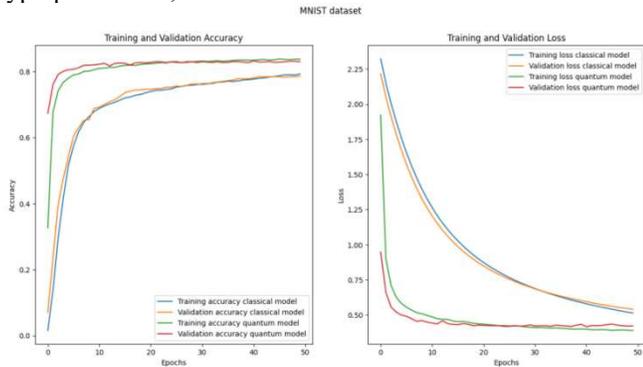

Fig. 7. Comparing a quantum classifier with an ideal simulator on the MNIST dataset against a classical classifier

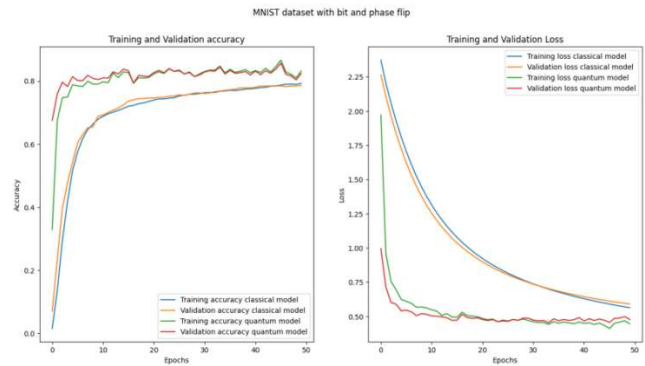

Fig. 8. Using the MNIST dataset, a quantum classifier with bit and phase flip errors is compared against a classical classifier

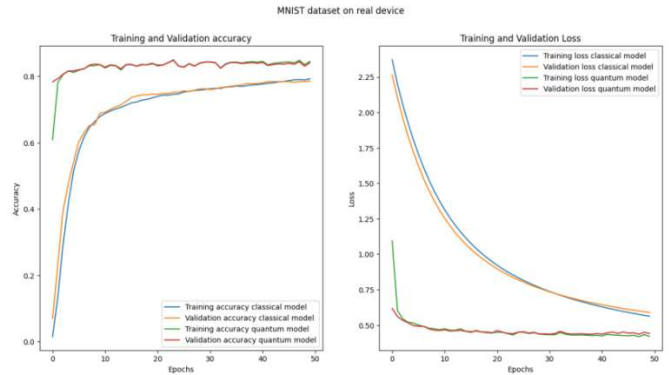

Fig. 9. When faced with a classical classifier, a quantum classifier trained on genuine quantum devices using MNIST datasets

TABLE I
DETAILS OF THE COMPARISON BETWEEN THE CLASSICAL CNN AND THE QUANTUM QCNN FOR MNIST DATASET

| Models | Parameters | Learning Rate | Qubits | Reuploading Layers | Flip Parameters | Test accuracy |
|---|---|---|---|---|---|---|
| CNN | ≈2900 | $10^{-3}$ | | - | - | 0.76 |
| QCNN (ideal) | ≈2800 | $10^{-3}$ | 5 | 5 | - | 0.82 |
| QCNN (noisy) | ≈2800 | $10^{-3}$ | 5 | 5 | 0.1 | 0.80 |
| QCNN (real) | ≈2800 | $10^{-3}$ | 5 | 5 | 0.1 | 0.81 |

Results indicate that the QNN outperforms the classical CNN in terms of accuracy and convergence time across all datasets. Notably, despite the increased complexity of the Fashion MNIST (refer to Figs. 10, 11, and 12) and CIFAR10 (refer to Figs. 13, 14, and 15) dataset, the QNN demonstrates advantages over the CNN, particularly in loss descent and overall accuracy. The hybrid approach, integrating classical TensorFlow convolutional layers with quantum layers, is employed for CDOPA due to its high-dimensional nature in Figs. 16, 17, and 18.

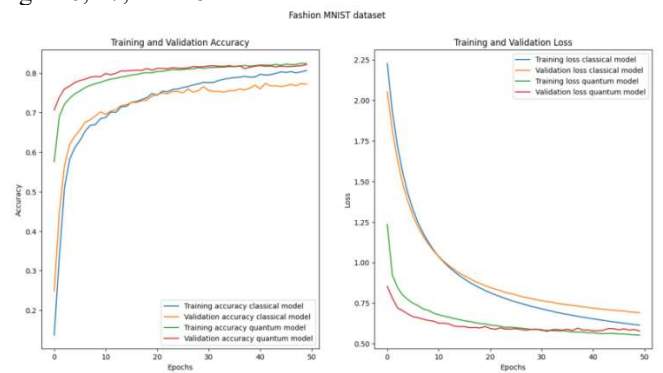

Fig. 10. Comparing a quantum classifier with an ideal simulator on the Fashion MNIST dataset against a classical classifier

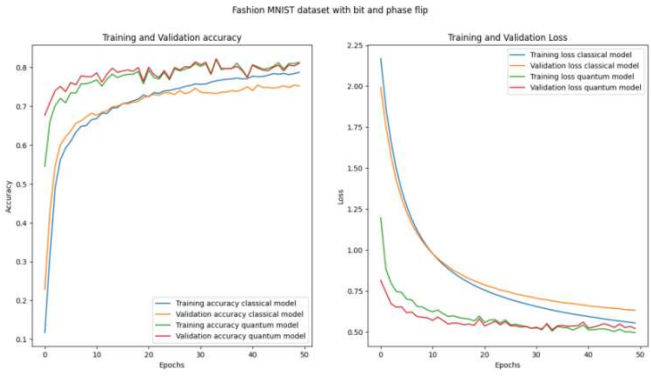

Fig. 11. Using the Fashion MNIST dataset, a quantum classifier with bit and phase flip errors is compared against a classical classifier

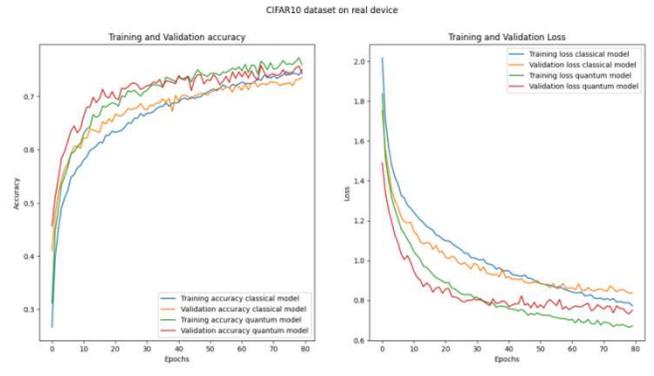

Fig. 15. When faced with a classical classifier, a quantum classifier trained on genuine quantum devices using CIFAR10 datasets

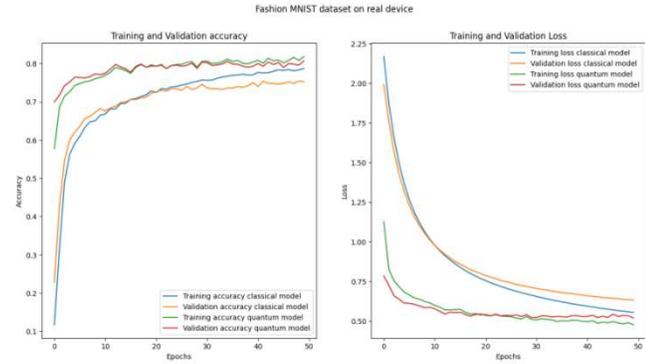

Fig. 12. When faced with a classical classifier, a quantum classifier trained on genuine quantum devices using Fashion MNIST datasets

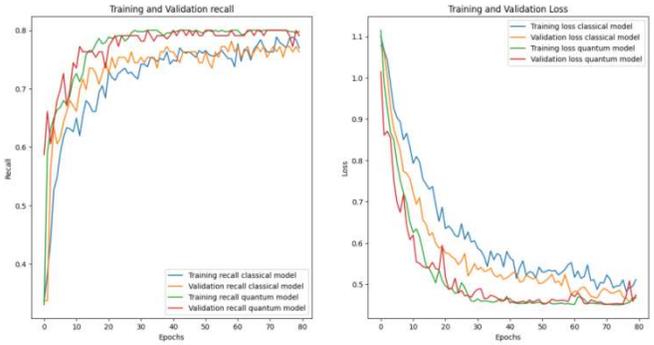

Fig. 16. Comparing a quantum classifier with an ideal simulator on the CDOPA dataset against a classical classifier

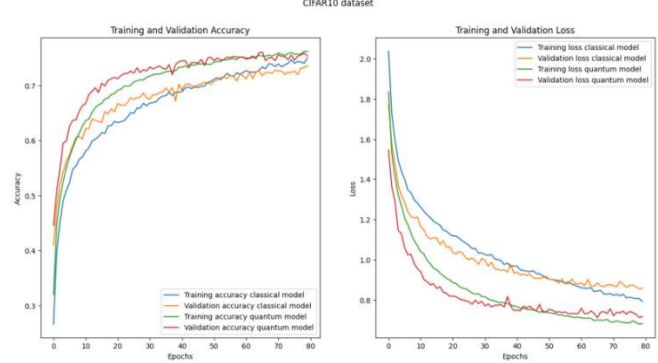

Fig. 13. Comparing a quantum classifier with an ideal simulator on the CIFAR10 dataset against a classical classifier

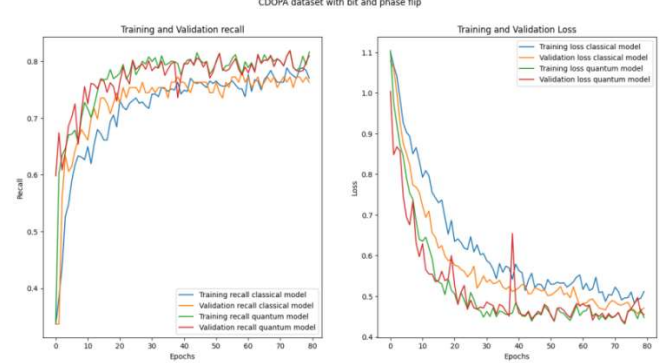

Fig. 17. Using the CDOPA dataset, a quantum classifier with bit and phase flip errors is compared against a classical classifier

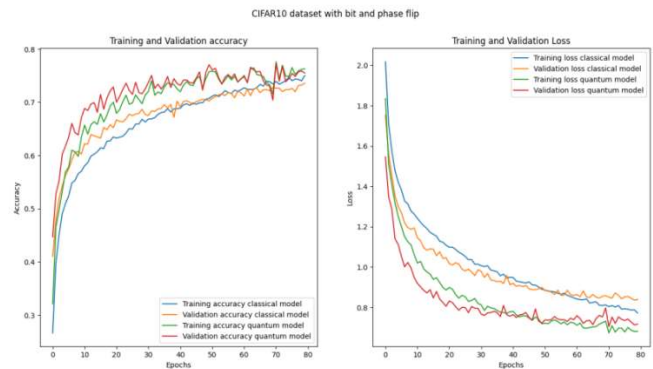

Fig. 14. Using the CIFAR10 dataset, a quantum classifier with bit and phase flip errors is compared against a classical classifier

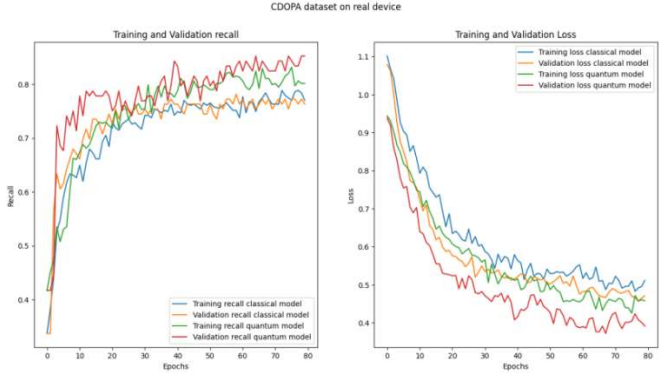

Fig. 18. When faced with a classical classifier, a quantum classifier trained on genuine quantum devices using CDOPA datasets

## B. Image Generation

As the exploration advances in quantum ML, another significant challenge in computer vision arises—image generation. This section presents findings obtained through the utilization of the patch GAN algorithm. Notably, the study employs three renowned computer vision benchmark datasets—MNIST, Fashion MNIST, and CIFAR10. The experimental setup comprises a discriminator-generator framework, featuring a conventional fully-connected discriminator paired with both classical and quantum generator models. The classical generator model adheres to a standard conditional convolutional generative network, whereas the quantum patch architecture is adapted to incorporate class labels, facilitating the generation of class-specific images. Due to the inherently qualitative nature of image generation tasks, human evaluation remains pivotal. A comparative analysis among original images, classical GAN-generated images, and quantum GAN-generated images is conducted to gauge the quantum model's efficacy. Furthermore, the training dynamics, inferred from the discriminator's loss value, are visualized as shown in Figs. 19, 20, and 21.

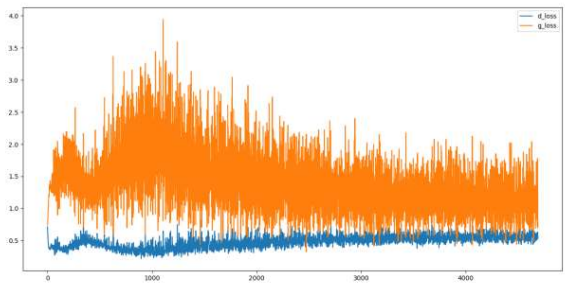

Fig. 19. Loss vs iteration steps of discriminator (blue) vs quantum generative (orange) model for MNIST dataset

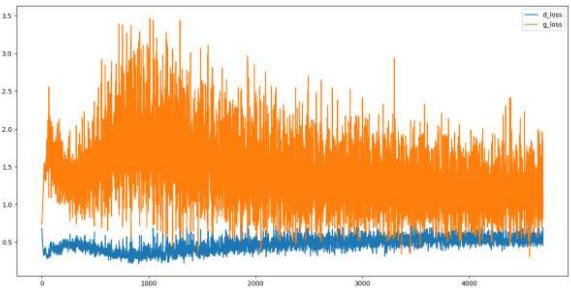

Fig. 20. Loss vs iteration steps of discriminator (blue) vs quantum generative (orange) model for Fashion MNIST dataset

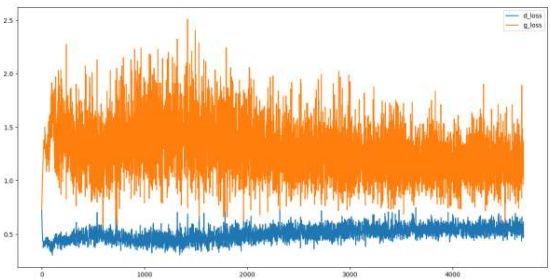

Fig. 21. Loss vs iteration steps of discriminator (blue) vs quantum generative (orange) model for CIFAR10 dataset

Similar to the development of the quantum classifier, the quantum generative model is seamlessly integrated into the PyTorch DL framework. Implemented as Python classes leveraging the Pennylane[24] library, this integration ensures compatibility and ease of use within the research framework. For the three datasets, the patch GAN has effectively replicated the images to the same standard as conventional classical generative models. Observing the results, it's evident that the quantum generative model exhibits comparable performance in reproducing both greyscale and RGB images without any anomalies during training. Specifically, the discriminator loss stabilizes at 0.5 (illustrated by the blue curves), indicating random guessing of generated image authenticity, while the quantum generator loss steadily decreases with each training step. The classical discriminator architecture comprises a Multi-layer Perceptron (MLP) with two hidden layers, consisting of 64 and 32 nodes, respectively. Both the generator's and discriminator's optimizers were configured with a learning rate of $2 \times 10^{-4}$. In comparison, the classical generator utilized is a standard convolutional model featuring two transposed convolution layers. Commencing from a latent space of dimension 50, this generator outputs the image, encompassing approximately 10,000 parameters. Conversely, for the quantum model, the considerably fewer qubits required result in a total trainable parameter count of approximately 4,500, encompassing 14 distinct sub-generators' circuits. Thus, despite a smaller quantity of trainable parameters, the quantum model demonstrates performance comparable to its classical counterpart.

## VI. CONCLUSION AND FUTURE WORKS

This study suggests that quantum computation holds promise for advancing ML, extending its applicability beyond physical research domains. Notably, experiments demonstrate that our image classifier achieves performance levels comparable to, if not surpassing, those of conventional classical architectures. Furthermore, our image generator faithfully reproduces benchmark datasets for image generation tasks. The utilization of simple yet hybrid VQCs has proven effective in mitigating bit and phase flip-like noise, particularly in low concentrations. However, it has become evident that amplitude damping errors significantly impede the performance of quantum algorithms. Given the current limitations of quantum devices, hybrid algorithms that leverage both quantum and classical computation present an intriguing avenue for exploration, potentially leading to quantum advantage. Future research should focus on refining embedding ansatzes, which play a crucial role in determining the feasibility of achieving quantum advantage. Moreover, the accessibility of quantum computers is paramount for assessing algorithm performance. As circuit complexity increases, simulators face exponential slowdowns, rendering even simple algorithms infeasible to execute. Real quantum computers offer the opportunity to evaluate algorithm performance authentically, circumventing the challenges posed by simulated noise. Addressing the computational demands of QNNs training requires innovative solutions. Libraries facilitating multiple executions of QNNs on the same quantum device are urgently needed, mirroring the efficiency of classical neural network training routines. Ultimately, realizing a genuine quantum advantage may necessitate the complete implementation of QNNs as quantum algorithms. However, the current generation of NISQ imposes constraints, requiring classical optimization routines and lacking the capacity for parallelized computations within large quantum circuits, thereby limiting the exploitation of quantum superposition and entanglement.

---

[24] https://pennylane.ai/

## VII. DECLARATIONS

*A.* **Funding:** No funds, grants, or other support was received.

*B.* **Conflict of Interest:** The authors declare that they have no known competing for financial interests or personal relationships that could have appeared to influence the work reported in this paper.

*C.* **Data Availability:** Data will be made on reasonable request.

*D.* **Code Availability:** Code will be made on reasonable request.